\documentclass[twocolumn,epsfig,floats,showpacs]{revtex4}
\usepackage{graphicx}
\usepackage{color}

\usepackage{fancyhdr}
\fancypagestyle{titlepage}{
\fancyhead[R]{J. Phys. Chem. C \textbf{126}, 21101-21108 (2022)}}
\begin{document}

\title{Rhombohedral and Turbostratic Boron Nitride Polytypes Investigated by X-ray Absorption Spectroscopy}

\author{Weine Olovsson}
\affiliation{Department of Physics, Chemistry and Biology (IFM), Link\"{o}ping University, SE-581 83 Link\"{o}ping, Sweden.}
\email{weine.olovsson@liu.se}
\author{Martin Magnuson}
\affiliation{Department of Physics, Chemistry and Biology (IFM), Link\"{o}ping University, SE-581 83 Link\"{o}ping, Sweden.}

\date{\today}


\begin{abstract}
The electronic structure of rhombohedral sp$^2$ hybridized boron nitride (r-BN) is characterized by X-ray absorption near-edge structure spectroscopy. Measurements are performed at the boron and nitrogen K-edges (1s) and interpreted with {\it first principles} density functional theory calculations, including final state effects by applying a core-hole.
We show that it is possible to distinguish between different 2D planar polytypes such as rhombohedral, twinned rhombohedral, hexagonal and turbostratic BN by the difference in chemical shifts. 
In particular, the chemical shift at the B 1s-edge is shown to be significant for the turbostratic polytype.
This implies that the band gap can be tuned by a superposition of different polytypes and stacking of lattice planes.
\end{abstract}

\maketitle

\section{Introduction}
Recently, the interest in boron nitrides has grown significantly due to the potential application in novel devices, e.g.\ in microelectronics for various optical devices in the UV range~\cite{Watanabe2009,Kudrawiec2020,Gil2022}, detectors~\cite{Uher2007} as well as in graphene based devices~\cite{Dean2010,Britnell2012}.
The reason is due to the desirable properties of wide band gap semiconductors with high thermal stability and mechanical strength~\cite{Magnuson2022}.
Moreover, for 2D layered BN with sp$^2$-hybridized in-planar bonding, there exist several different polytypes of crystalline structures; the hexagonal (h-BN)~\cite{Pease1950}, the
rhombohedral (r-BN)~\cite{Ishii1981} and its twinned version (twin-r-BN)~\cite{Oku2003}, and a less ordered turbostratic phase (t-BN)~\cite{Thomas1963}. While there are numerous studies in the literature of the cubic, wurtzite and hexagonal BN polytypes, there are relatively few attempts to address the rhombohedral phase and no calculations of its X-ray absorption near-edge structure (XANES) spectra.
Only recently, it has been possible to produce high-quality r-BN films by using chemical vapor deposition (CVD)~\cite{Chubarov2012}. 
In contrast with the ordered phases, t-BN constitutes a random-layer lattice in which the planes are stacked roughly parallel to each other, along the plane normal, with a random rotation and random translation~\cite{Thomas1963}.

The electronic structure of layered sp$^2$ BN are characterized by strong in-plane $\sigma$-bonding in the hexagonal lattice and weaker $\pi$-bonds between the planes. This gives rise to prominent features in the K-edge (1s) absorption spectra, a sharp $\pi^*$ resonance at the core-edge and a broader $\sigma^*$ peak in the higher energy region at 6-8 eV from the edge.
These features are especially pronounced at the B sites and less prominent for N.
Recently, a splitting of 0.10 eV found at the $\pi^*$ B K-edge in h-BN~\cite{Li2012} was interpreted as due to a dislocation of planes causing a different stacking sequence~\cite{McDougall2014}.
While many theoretical XANES assessments rely on density functional theory (DFT)~\cite{Kohn1999} supercell techniques, other methods such as the many-body perturbation theory in the form of the Bethe-Salpeter equation (BSE) can also be utilized~\cite{Carlisle1999,Shirley2000, Olovsson2019}.
The sharp B $\pi^*$ peak in h-BN is attributed to a single strongly bound core-exciton as previously observed in BSE calculations~\cite{Olovsson2019}. The peak splitting can be attributed either to a double phase structure or a sign of defects in the material.

Firstly, a correct determination of the crystalline structure should be addressed. As the in-plane lattice constant and the spacing between the basal planes are almost identical for h- and r-BN, differing with a variation in the stacking~\cite{Li2003}, most experimental tools are not suitable to distinguish between them. While it is straightforward to pick out planar and cubic crystal structures using e.g.\ X-ray diffraction (XRD), it is more challenging to separate between different sp$^2$ polytypes due to their similarity. However, pole figures in XRD have been useful to characterize the structure of r-BN~\cite{Chubarov2011,Chubarov2012,Chubarov2013,Chubarov2014,Chubarov2015,Chubarov2018}. 

Secondly, in order to tune the band gap properties of BNs, as well as for understanding their underlying chemical bonding and electronic structure, it is vital to be able to distinguish between different phases by using reliable experimental and theoretical techniques. 
XANES spectroscopy is an element selective method, probing the unoccupied states in a material. Due to the sensitivity of the spectra to the surrounding chemical environment of an atom, it is a powerful technique for structural characterization. 

In the present work, we employ high-resolution XANES measurements at the B and N K-edges (1s) obtained by using synchrotron radiation, combined with {\it ab initio} calculations within the highly accurate full potential augmented plane wave and local orbitals (APW+lo) scheme~\cite{Sjostedt,wien2k,wien2k-2} for DFT. By computing the spectrum of different BN polytypes, we show that it is possible to distinguish between their characteristic peak structures and energy shifts. 
We model the turbostratic phase by including random {\it translations} of planes, while all random {\it rotations} are neglected.  
In this way, our t-BN models preserve an important aspect of randomness, while supercell sizes are not a limiting factor in the calculations.

\section{Methodology}

\subsection{Theory}
The calculated XANES spectra were obtained by {\it first principles} DFT~\cite{Kohn1999} methodology.
In more detail, the all-electron full potential APW+lo scheme~\cite{Sjostedt}, as implemented in the {\tt WIEN2k} software package~\cite{wien2k,wien2k-2}, was utilized.
In order to account for the final state effect, a 1s core-electron at a single atom site was promoted to the valence band, leaving a core-hole behind.
Thereafter, the transition rate between the core-level and the unoccupied density of states at the ionized atom was computed according to Fermi's golden rule 
and within the electric dipole approximation ($\Delta l \pm 1$).
In this work, the exchange-correlation function is set to the generalized gradient approximation (GGA) according to Perdew et al.~\cite{PBE1996}.
For a brief overview of different methodologies to obtain absorption spectra, see e.g.\ ref~\cite{Mizoguchi2010} and references within.

Since no significant difference between the spectra obtained by BSE and supercell techniques was found in the case of h-BN~\cite{Karsai2018}, 
we deem that the DFT supercell technique utilizing the core-hole approximation is adequate for a comparative study of layered BN polytypes.
Furthermore, due to the close similarities between the sp$^2$ BN structures, we assume that corrections to the band structure such as provided by the GW approximation~\cite{Aryasetiawan1998}, or by using other schemes, will not significantly affect the comparison.

All the layered boron nitride structures consist of 2D hexagonal lattice planes with B and N atoms, where the other element constitutes the nearest neighbors.
In this work, we consider the rhombohedral, twinned rhombohedral and hexagonal structures, as well as the planar disordered turbostratic structure.
For comparison, we also contrast the results with that of cubic BN and a single monolayer (ML) of hexagonal BN.
Experimental lattice parameters were used in the calculations (see Table 1).
The main difference between the 2D planar phases are in the stacking of the layers.
This is illustrated in Fig.\ 1, showing the AB... stacking in h-BN and ABC... in r-BN.
Twinned r-BN has an ABCA'B'C'... stacking, which can be described as follows; by taking the unit cell of the r-BN structure, doubling it along the c-axis and rotating 
the repeated cell by 60 degrees~\cite{Chubarov2018}.
Note that while AB does not refer to the same stacking between h- and r-BN, ABC is the same between r- and twin-r-BN. 
In the turbostratic phase there is no preferred orientation of the planes in the different layers, i.e.\ it is random.
Here, the spacing in-between the planes, d, is slightly increased by $\sim$0.05 \AA\ for the present sample, with a larger variation compared with the other structures.

To model the turbostratic BN structure is a challenge due to the inherent random orientation of each plane, meaning that a unit cell can be arbitrarily large.
Therefore, we model t-BN by starting from the cell of a hexagonal ML and then apply a fixed random translation to the B and N atom pair in each different plane.
This means that the t-BN unit cell will be kept minimal with only 2 atoms in each plane, since random plane rotations are not included.
In this way, the challenging case of a purely turbostratic structure of arbitrarily large size is approximated by a computationally feasible cell, keeping an important part of the randomness.
In the literature, t-BN has also been modelled without considering randomness of plane translations and rotations, instead stacking planes from ordered structures randomly~\cite{Mengle2019}.

For computing the XANES spectra, supercells are chosen with 3 x 3 planes containing 18 atoms in order to exclude spurious interaction among core-ionized sites due to periodic boundary conditions. The utilized supercells have 3 layers for r-BN (54 atoms), 4 for h-BN (72 atoms), 6 for twin-r-BN (108 atoms) and 7 for t-BN (126 atoms). The single ML BN system thus has 18 atoms, while an effective c-axis of 15 \AA\ was chosen. For t-BN, in total ten different model structures were generated, denoted from "t1-BN" to "t10-BN". An example of two of the structures, t5-BN and t6-BN are shown in Fig.\ 2. Finally, a Lorentzian broadening of 0.2 and 0.4 eV full width at half maximum (FWHM) was applied to the B K-edge and N K-edge spectra, respectively, taking the effect of the core-hole lifetime into account.

\begin{figure}
\includegraphics[width=70mm]{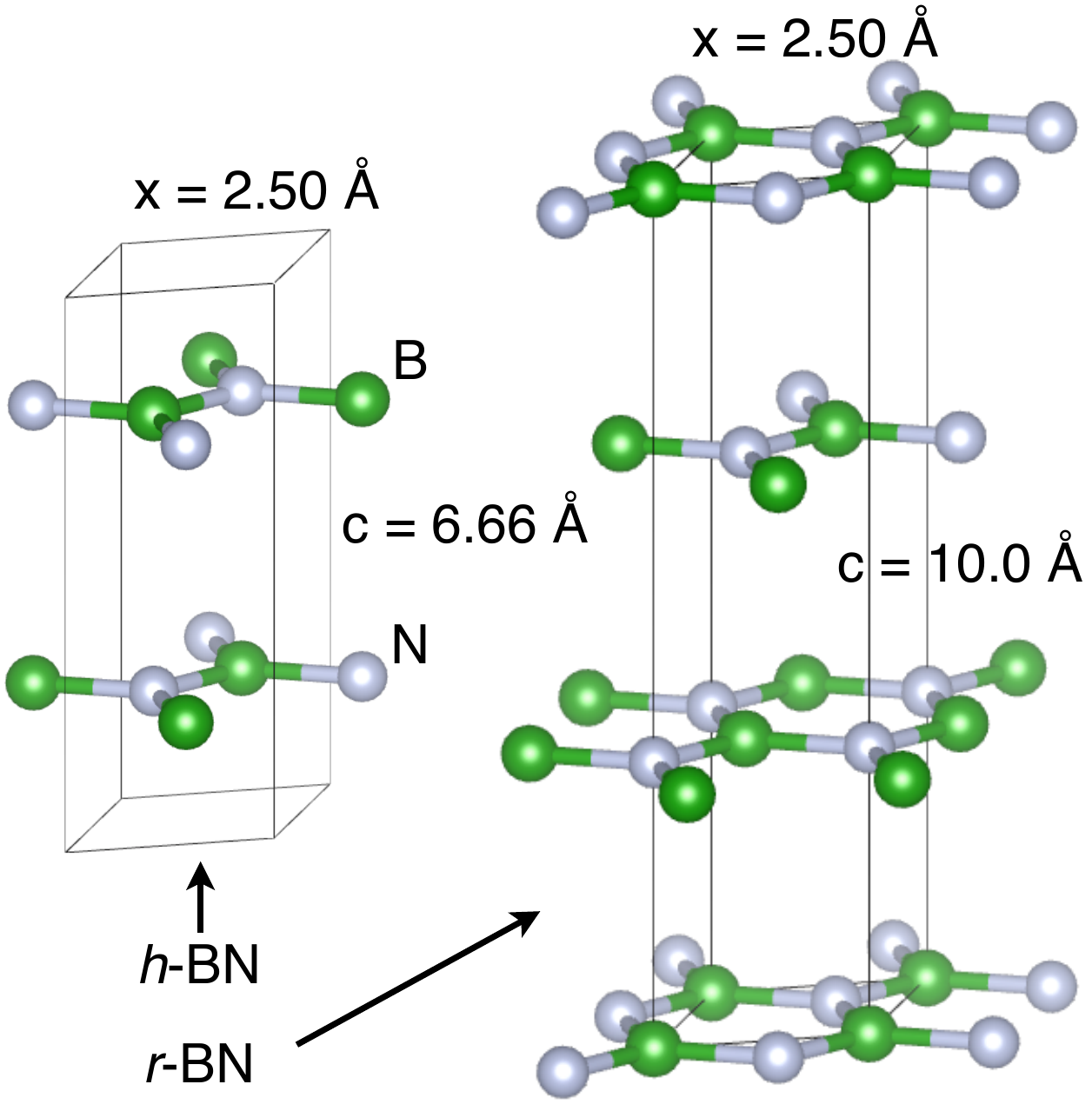} 
\vspace{0.2cm} 
\caption[] {The AB respectively ABC stacking in the hexagonal (left) and rhombohedral (right) BN polytypes. 
The figure was made with help of {\tt VESTA}~\cite{vesta}.
} 
\label{fig1}
\end{figure}

\begin{figure}
\includegraphics[width=70mm]{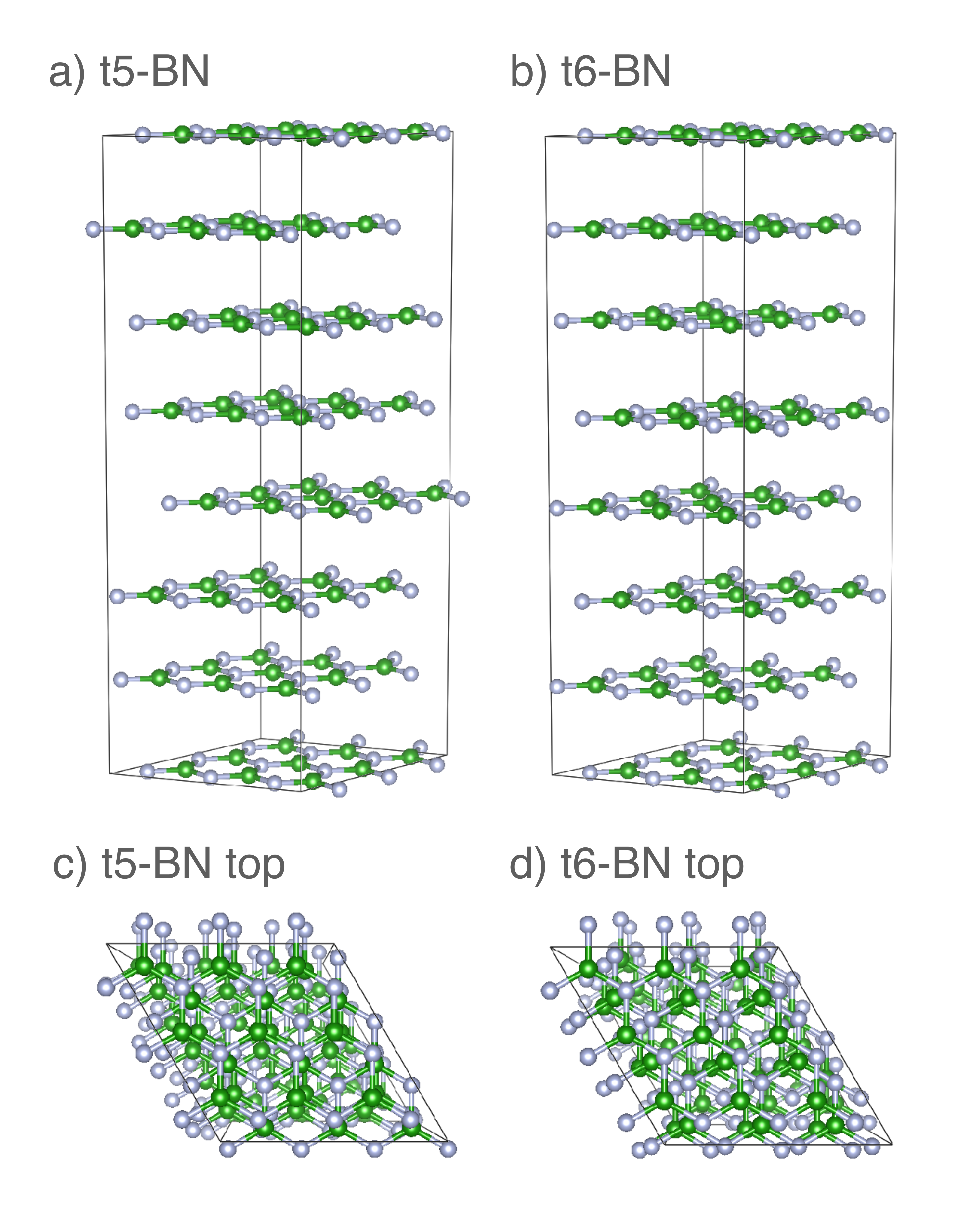} 
\vspace{0.2cm} 
\caption[] {Two different turbostratic BN model supercells, denoted as "t5-BN" and "t6-BN", with 3 x 3 planes and 7 layers with a random translation. 
A side view of the a) t5-BN and b) t6-BN supercells are shown, together with a view from the top for c) t5-BN and d) t6-BN.
The figure was made with help of {\tt VESTA}~\cite{vesta}.
} 
\label{fig2}
\end{figure}

\subsection{Measurements}
Soft X-ray absorption near-edge structure measurements were performed at the undulator beamline I511-3 at MAX II (MAX-lab Laboratory, Lund, Sweden), comprising a 49-pole undulator, and a modified SX-700 plane grating monochromator~\cite{Magnuson1}. The measurements were performed with a base pressure lower than 5$\times$10$^{-7}$ Pa. In order to enhance the $\pi^*$ contribution that is suppressed at normal incidence~\cite{Magnuson2}, the XANES spectra were measured at 20$^{\circ}$ grazing incidence angle in total fluorescence yield (TFY) mode. The energy resolutions were 0.025 and 0.1 eV, at the B 1s and N 1s absorption edges, respectively. The XANES spectra were normalized to the step before, and after the absorption edges.

\subsection{Sample growth}
A hot wall CVD reactor with a SiC coated graphite susceptor was utilized for the deposition of sp$^2$-BN thin films. 
As a substrate, c-axis oriented $\alpha$-Al$_2$O$_3$ was employed. To reduce the lattice mismatch between sapphire and hexagonal boron nitride (4.67 \AA\ and 2.51 \AA\ respectively), 
an aluminum nitride (lattice parameter 3.11 \AA) buffer layer was formed by nitridation of the $\alpha$-Al$_2$O$_3$ surface at the growth temperature by introducing ammonia in the reactor with a concentration of 9.6\% 
for 10 minutes prior to a growth~\cite{Chubarov2011}.
The growth pressure was 100 mbar and H$_2$ was used as carrier gas. As boron and nitrogen precursor, triethyl boron (TEB) and ammonia (NH$_3$) were used, respectively. 
After substrate surface nitridation, the concentration of ammonia in the reactor cell was adjusted to 6\% in order to achieve the desired nitrogen to boron ratio of 614~\cite{Chubarov2012}.
The growth temperature was 1200 $^\circ$C. As the source of the Si for the stabilization of crystalline sp$^2$-BN growth, a SiC protective coating of the susceptor was utilized.
For the initial investigation of the grown films, X-ray diffraction in Bragg-Brentano geometry was conducted to evaluate crystalline quality of the deposited layers,
employing a Philips powder diffractometer equipped with Cu anode X-ray tube using K$_{\alpha 1,2}$ radiation and filtering Cu K$_\beta$ radiation by Ni filter.

\section{Results and discussion}

\begin{figure}
\includegraphics[width=90mm]{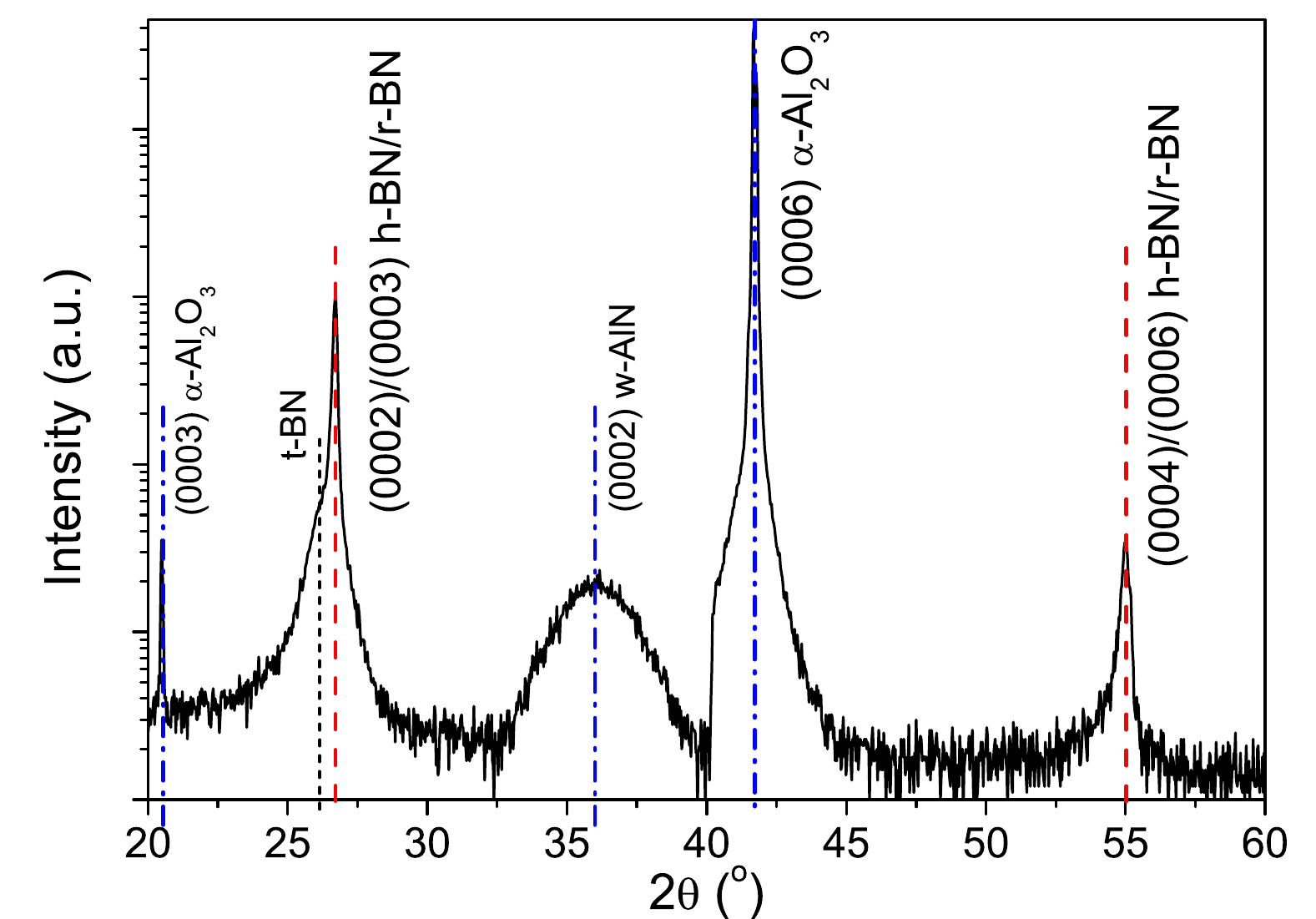} 
\vspace{0.2cm} 
\caption[] {XRD pattern of crystalline sp$^2$-BN deposited at the conditions described in the main text.} 
\label{fig3}
\end{figure}

Figure 3 shows XRD data recorded from a r-BN sample deposited by CVD as described in the previous section. 
The data are dominated by two peaks that correspond to crystalline sp$^2$-BN located at 26.7$^\circ$ and 55.0$^\circ$ in 2$\theta$ that correspond to (0002) or (0003) planes of h-BN or r-BN 
and second order diffraction from these planes (red dashed lines in Fig.\ 3), respectively~\cite{jcpds}. Blue dash-dotted lines mark peaks associated with the $\alpha$-Al$_2$O$_3$ substrate and the AlN buffer layer.
The positions, sharpness and high intensity of these BN peaks indicate high crystalline quality of the sp$^2$-BN film. On the contrary, the width of the AlN buffer layer peak located at 36$^\circ$ is rather large, which is characteristic of a strained layer. 

The high intensity of the BN peaks at 26.7$^\circ$ and 55$^\circ$ suggests that the crystal structure of the layer is mainly rhombohedral.~\cite{jcpds}
However, there is also a shoulder on the low-energy side of the 26.7$^\circ$ crystalline sp$^2$-BN peak that is associated with the formation of less ordered crystalline sp$^2$-BN -- t-BN (black dotted line). 

From the XRD peak structures the r-BN reflection at 26.7$^\circ$ corresponds to a c-value of 10.00 \AA\ (interlayer spacing d = 3.33 \AA) and the t-BN reflection centered at 26.3$^\circ$ corresponds to a c-value of 10.15 \AA\ (d = 3.38 \AA), if including 3 planes as for the rhombohedral polytype. The t-BN structure has a width with a minimum angle of 25.0$^\circ$ corresponding to a maximum c = 10.67 \AA\ (d = 3.56 \AA) and a maximum angle at 28.0$^\circ$ corresponding to a minimum c = 9.55 \AA\ (d = 3.18 \AA).

Since the twinned version of r-BN was previously found in similar CVD deposited rhomobohedral samples from Chubarov et al.~\cite{Chubarov2011,Chubarov2012,Chubarov2013,Chubarov2014,Chubarov2015,Chubarov2018}, we also include it in the present study for comparison. In the theoretical calculations, it was simply modelled by using the measured lattice parameters of the r-BN structure, doubled along the c-axis.

\begin{table*}
\caption{Band gap and lattice parameters of BN polytypes}
\begin{tabular}{ccccccc}
\hline
$\textnormal{System}$&B$_{g}$ exp. (eV)&B$_{g}$ calc (eV)&$a$ ({\AA} )&$c$ ({\AA})&$d$ ({\AA}) \\
\hline
$\textnormal{twin-r-BN}$&$ - $&$3.83$&$2.504$&$20.016$&$3.336$\\
\hline
\hline
\end{tabular}
\label{table1}
\end{table*}

For reference, the band gaps and lattice parameters of different boron nitride polytypes are listed in Table 1. 
While all the experimental band gaps are from the literature, the theoretical values are obtained for the respective unit cells.
Here, one can note the wide spread in the measured band gaps for h-BN, between 3.6 - 7.1 eV. There are several reasons for this, related both to the specific methodology and quality of different h-BN samples~\cite{Solozhenko2001,Watanabe2004,Evans2008}. This includes the possibility of different stacking between the layers, see for instance ref~\cite{Lee2021} and references within.
A recent measurement by Yamada et al. gives 5.97 eV for the band gap in h-BN, while they find a smaller value, 5.82 eV, for t-BN~\cite{Yamada2021}.
In case of t-BN, c is shown for the model structures including 7 planes. The theoretical t-BN band gap varies between 3.76 - 3.96 eV, with the shown value taken as the average over the 10 t-BN models, which happens to coincide with t6-BN.
For the 1 ML BN system, the smaller and larger experimental values refer to the optical, respectively the electronic, band gap~\cite{Roman2021}.
The different 2D layered BN polytypes are compared to the cubic (zincblende) phase, c-BN.
It is well known that using the PBE GGA exchange-correlation function~\cite{PBE1996} typically underestimates the size of the band gaps in comparison with measurement. 
The calculated band gaps of h-, r- and c-BN agree well with previous ones computed within the same methodology~\cite{Ahmed2007}.

\begin{figure}
\includegraphics[width=90mm]{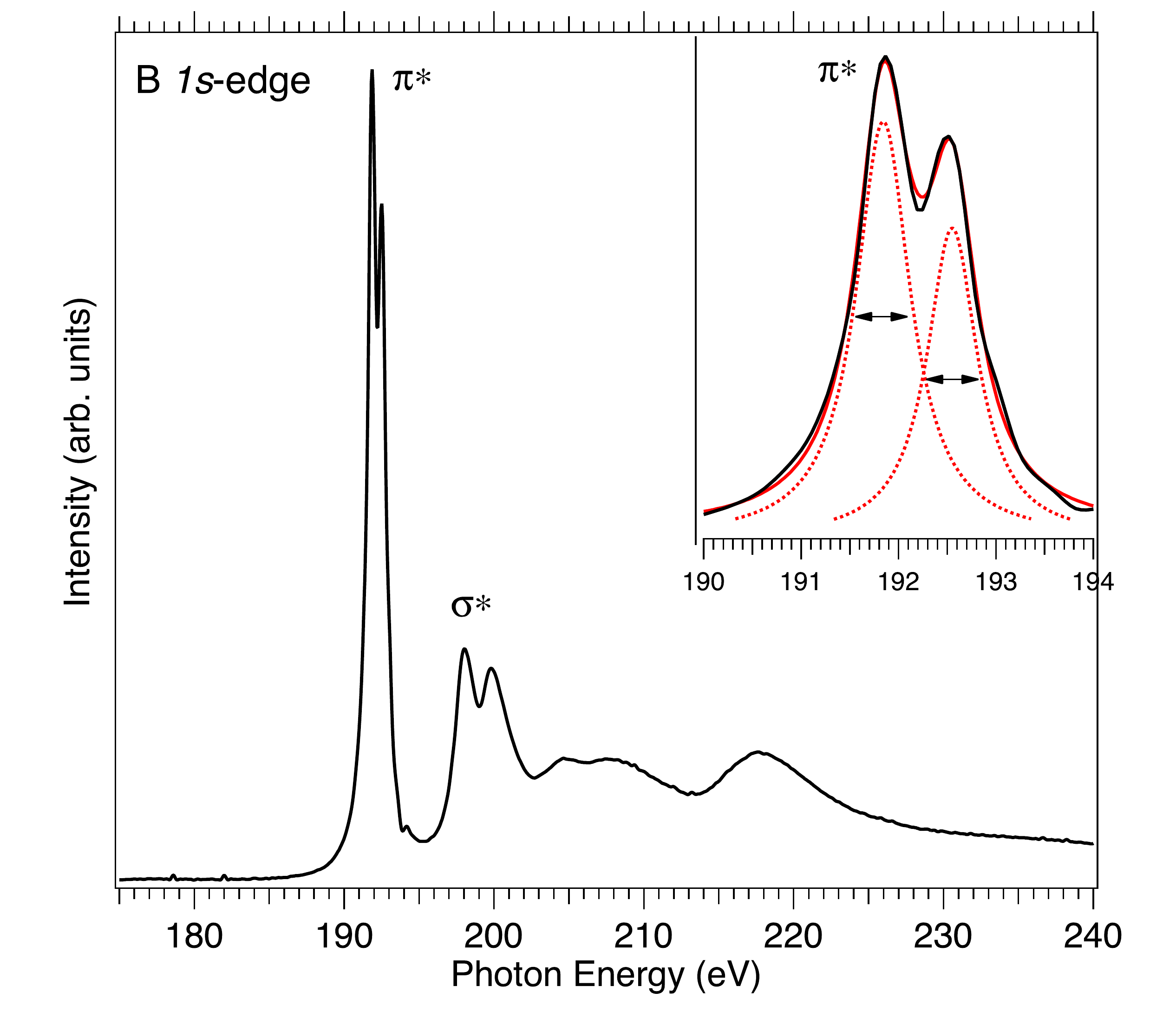} 
\vspace{0.2cm} 
\caption[] {Measured B K-edge XANES of the BN sample. The inset shows a close-up of the splitting at the $\pi^*$ peak.}
\label{fig4}
\end{figure}

Figure 4 shows the experimental XANES spectrum measured at the B K-edge (1s) of the sp$^2$ boron nitride sample. XANES spectra of layered BN typically exhibit a dominating $\pi^*$ core-edge, followed by a broader $\sigma^*$ region at 6-8 eV higher energy. Most prominently, we observe a splitting of 0.7 eV at the (191.84-192.55 eV) sharp $\pi^*$ peak, where the low-energy contribution is the most intense. The split is enlarged in the inset of Fig.\ 4 for clarity. 
At higher energy (198.0-199.8 eV), a broader double peak with 1.8 eV splitting is due to $\sigma^*$ states.  This is often referred to as a "camel-back" feature~\cite{Carlisle1999,Shirley2000} and was recently described as due to vibrational effects in h-BN~\cite{Karsai2018,Olovsson2019} present already at low temperature. The camel-back feature can be compared to a similar $\sigma^*$ region at the C K-edge in graphite~\cite{Olovsson2019}. At higher energies, the features are further broadened by multi-electron excitations. 

\begin{figure}
\includegraphics[width=90mm]{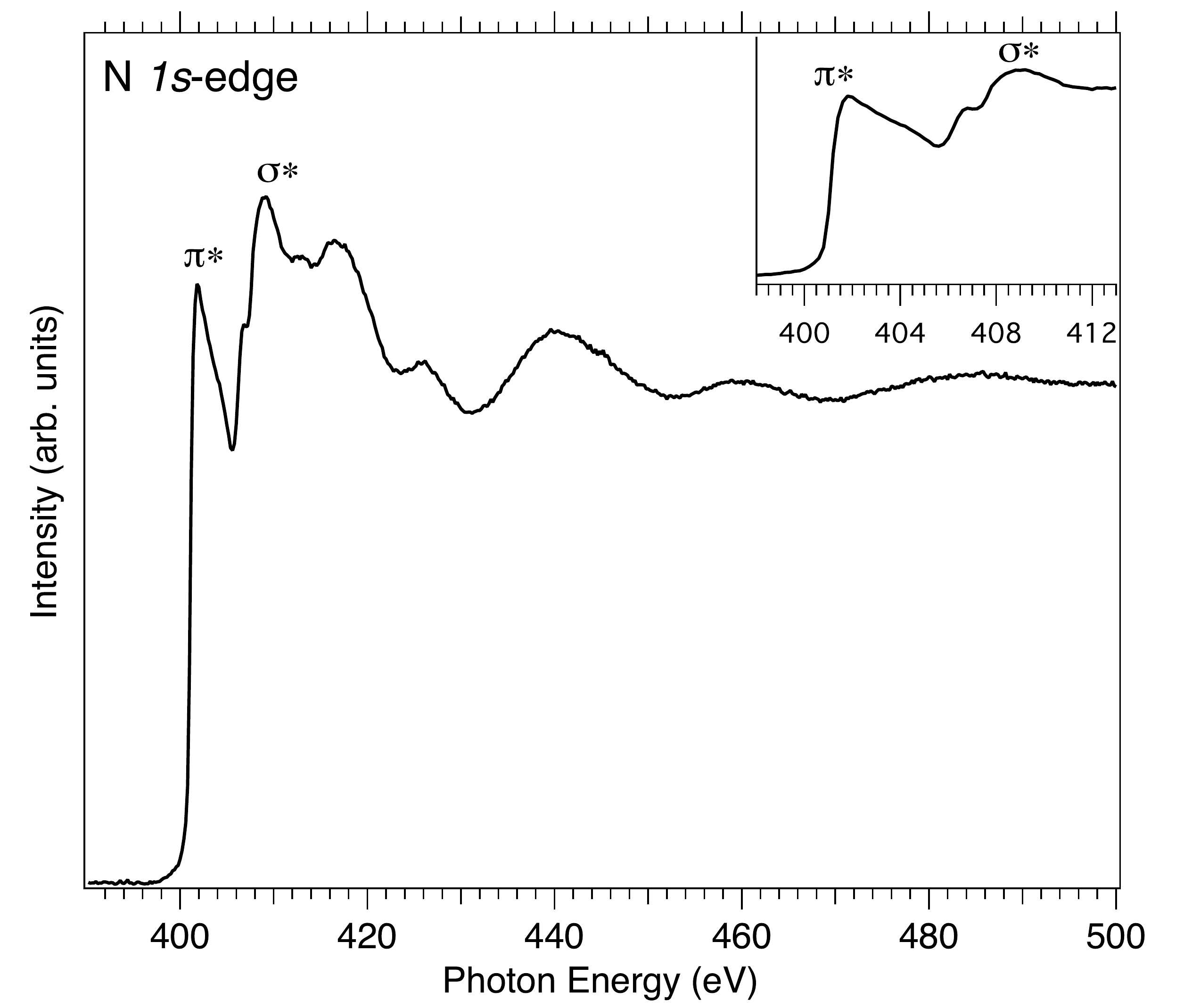} 
\vspace{0.2cm} 
\caption[] {N K-edge XANES measurements of the BN sample. The inset shows a close-up on the $\pi^*$ and $\sigma^*$ regions.}
\label{fig5}
\end{figure}

Figure 5 shows a N K-edge XANES spectrum of the same sample, where the inset is a close-up of the near-edge region. The $\pi^*$ core-edge peak at 401.8 eV has a sharp low-energy edge with a broad shoulder on the high-energy side. The $\sigma^*$ peak at 409 eV has a low-energy peak located at 406.8 eV. At higher energies, there are increasingly broad features due to multi-electron excitations. While the overall features of the B and N spectra in Fig. 4 and 5 are similar to the ones previously observed for other layered boron nitride systems, the most significant difference is the large splitting at the $\pi^*$ peak at the B K-edge. We are not aware of previous observations in the literature of such a large splitting in 2D planar BN polytypes.

\begin{figure}
\includegraphics[width=90mm]{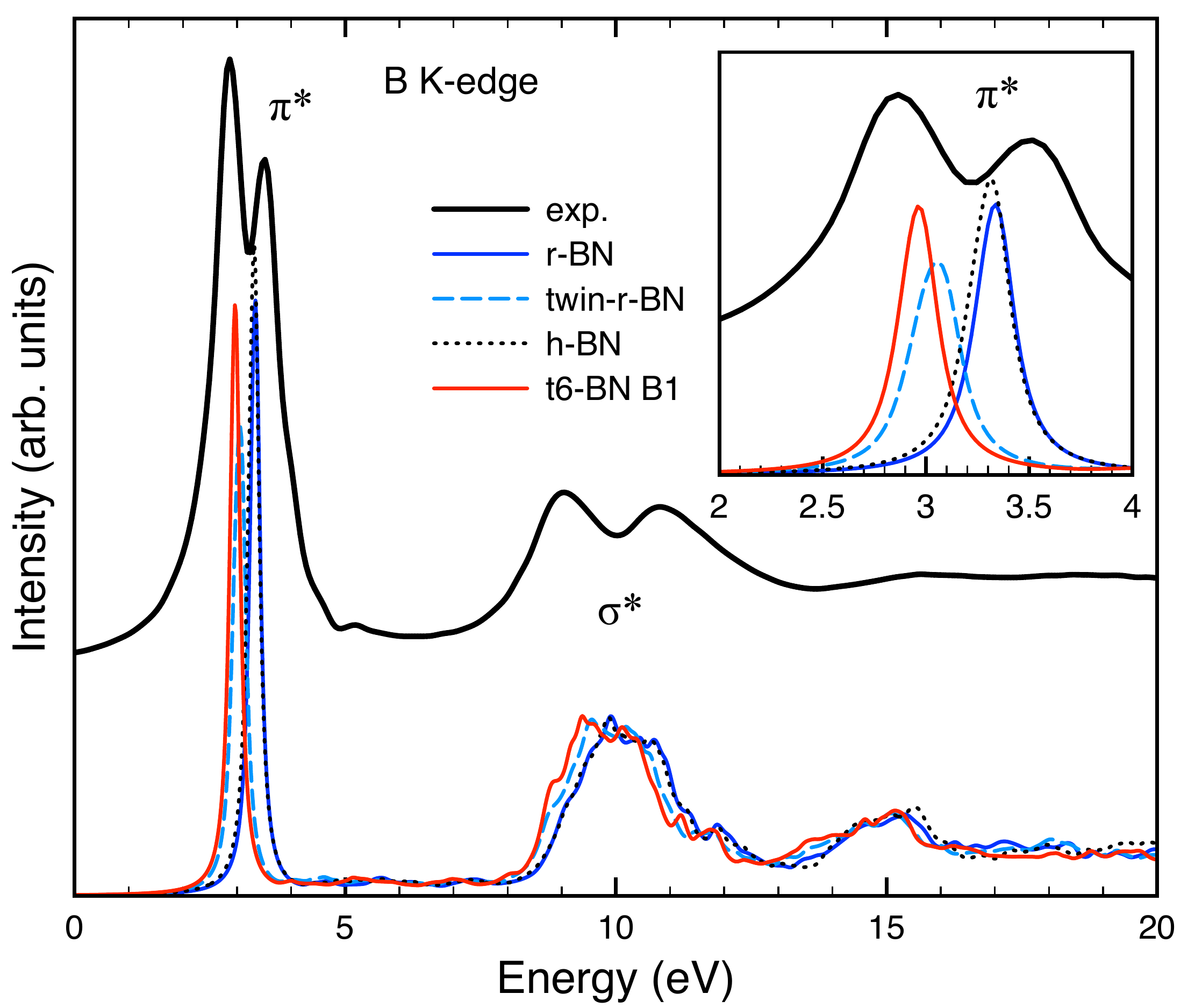} 
\vspace{0.2cm} 
\caption[] {Theoretical B K-edge XANES spectra of the rhombohedral (blue line), twinned rhombohedral (cyan dashed line), 
hexagonal (dotted line) and turbostratic (red line) BN phases, with comparison to measurement (lifted thick line).
The inset shows a close-up of the splitting at the $\pi^*$ peak.}
\label{fig6}
\end{figure}

\begin{figure}
\includegraphics[width=90mm]{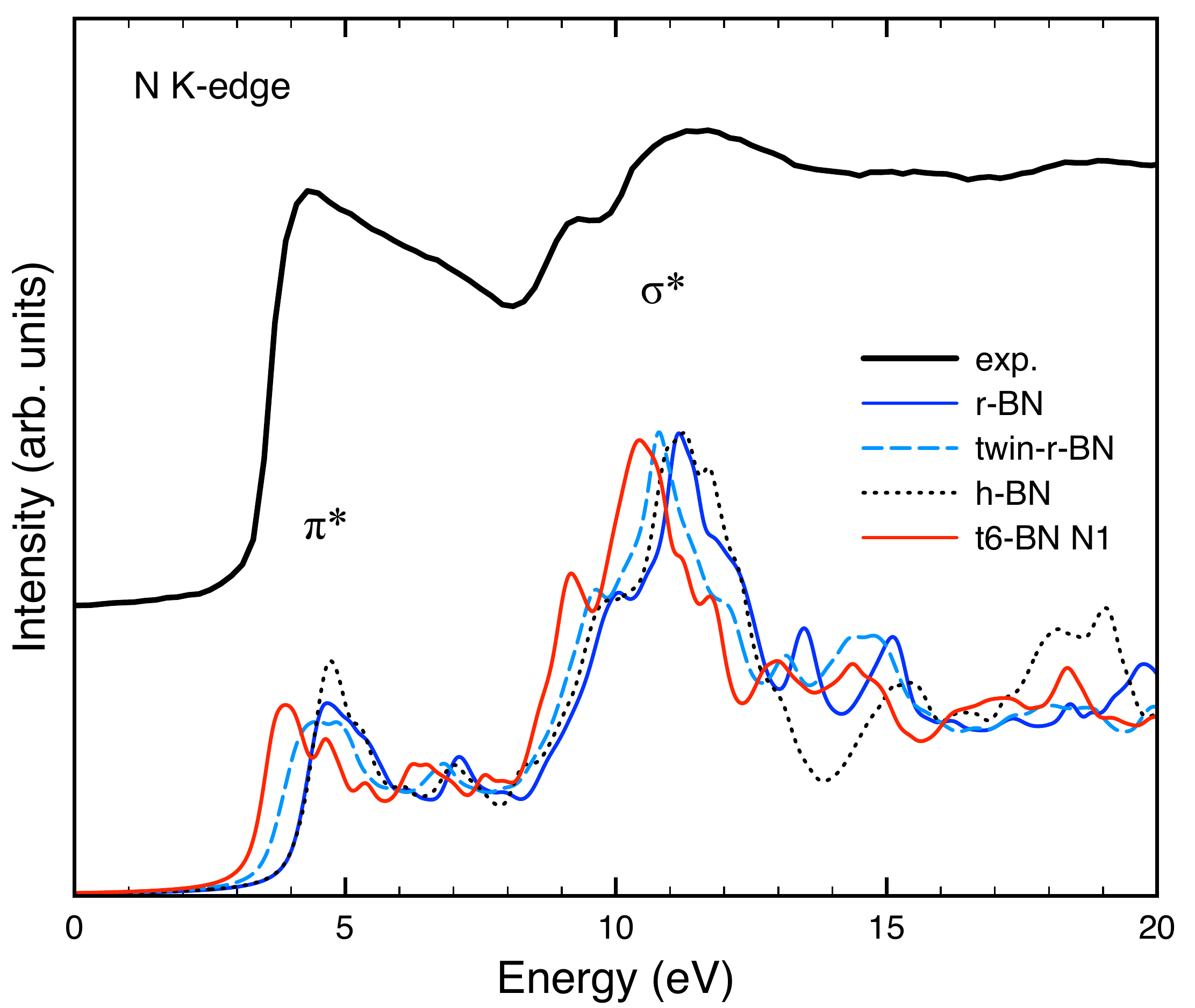} 
\vspace{0.2cm} 
\caption[] {Theoretical N K-edge XANES spectra of the rhombohedral (blue line), twinned rhombohedral (cyan dashed line),
hexagonal (dotted line) and turbostratic (red line) BN phases, with comparison to measurement (lifted thick line).}
\label{fig7}
\end{figure}

Figures 6 and 7 show the corresponding theoretical B and N spectra at the K-edge for the different layered BN polytypes; rhombohedral (blue line), twinned rhombohedral (cyan dashed line), hexagonal (dotted line) and turbostratic (red line) from the t6-BN model with atoms from the first plane from the bottom. As observed, the main features at the $\pi^*$ and $\sigma^*$ excitations are reproduced in the calculations. As discussed below, the $\pi^*$ splitting at the B K-edge can be reproduced from calculations by a superposition of at least two polytypes. At the N K-edge, it is not possible to distinguish a similar clear splitting in the measured spectrum. 

\begin{figure}
\includegraphics[width=90mm]{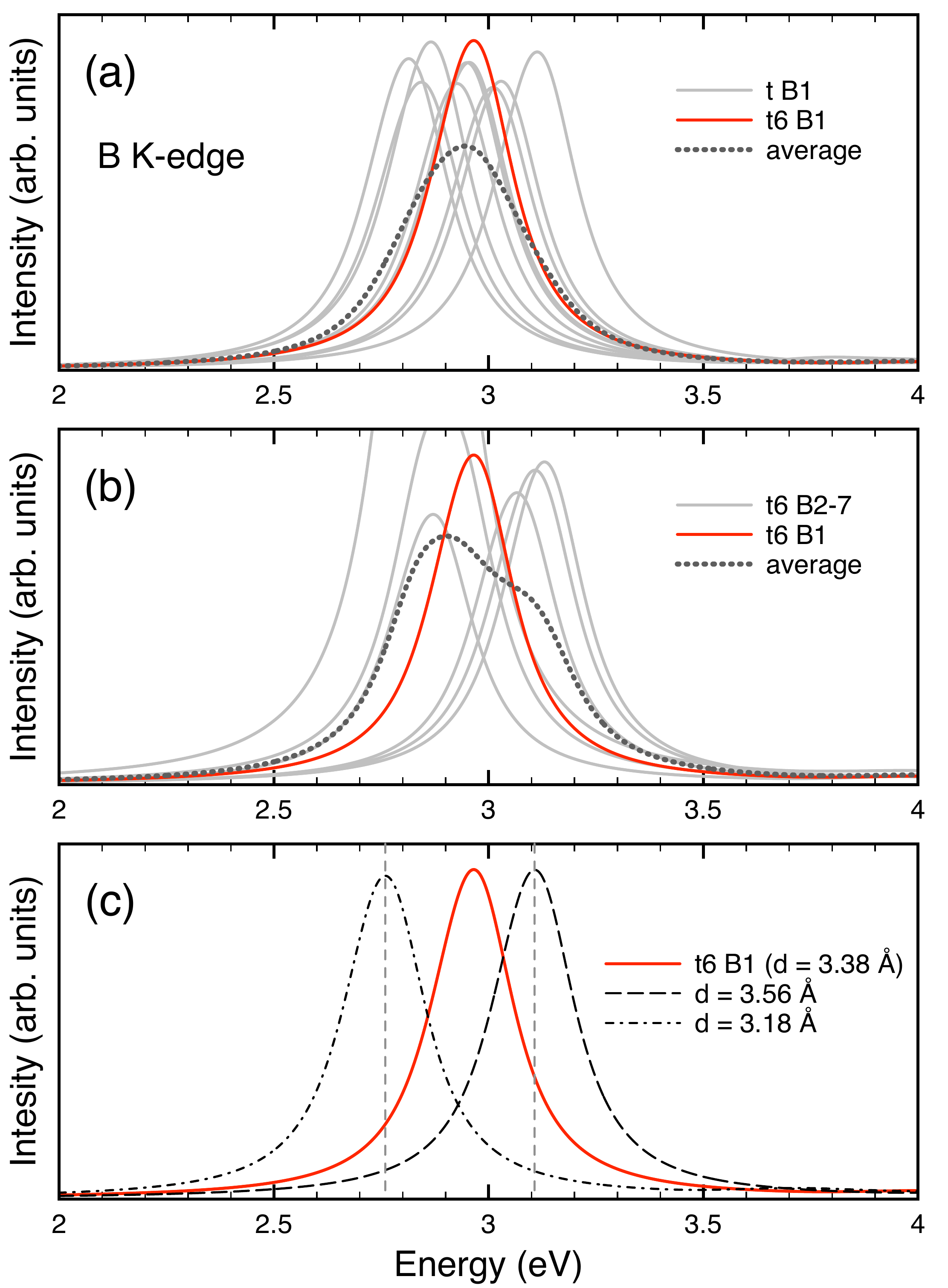} 
\vspace{0.2cm} 
\caption[] {Theoretical B K-edge XANES spectra for the 10 turbostratic model structures from 
a) atom B1 (gray lines), B1 in the model "t6-BN" (red line), and the average (dark gray dotted line), 
b) from the model "t6-BN", atoms on each plane (gray lines), B1 in the model "t6-BN" (red line), and the average (dark gray dotted line),
c) from the model "t6-BN", B1 (red line), B1 with a larger (dashed line) and a smaller (dot-dashed line) interplanar distance d.}
\label{fig8}
\end{figure}

In Figure 8 we summarize our results for the theoretical B K-edge XANES of the 10 turbostratic model systems, focusing on the $\pi^*$ peak. 
Here, the B atom at the first plane from the bottom, denoted as B1, is selected for computing a spectrum for each structure. Fig.\ 8 a) shows the dispersion over the different t-BN models (gray lines) compared with the t6-BN model (red line) and the average of the systems (dark gray dotted line), b) shows the results for the t6-BN model, with the different layers B2-7 (gray lines) compared with B1 (red line) and the average (dark gray dotted line), c) a comparison is made at B1 in t6-BN with the d interplanar distance (red line), with the limiting values from XRD, for the larger (dashed line) and smaller (dot-dashed line) distances. 
Here, an increased interplanar distance causes a shift towards higher energy.
Overall, the shown t-BN model $\pi^*$ peaks are roughly dispersed between 2.7 - 3.1 eV.
For all these model structures the effect of random rotations of the planes are not included, as discussed in the Methodology, Theory section.

Compared with the ordered phases, the interplanar distance has a large spread for the partially disordered turbostratic structure. 
In the literature, values or ranges has been found, for example, to be 3.35 - 3.50~\cite{Kobayashi2008}, 3.39~\cite{Coudurier2013} and up to 3.56 \AA~\cite{Rousseau2021}, compared with the present result of d = 3.38 (3.18 - 3.56) \AA. This gives an average interplanar distance which is overall $\sim$0.05 \AA\ larger than that of the ordered phases, as seen in Table 1, although it can also be distributed smaller.

\begin{figure}
\includegraphics[width=90mm]{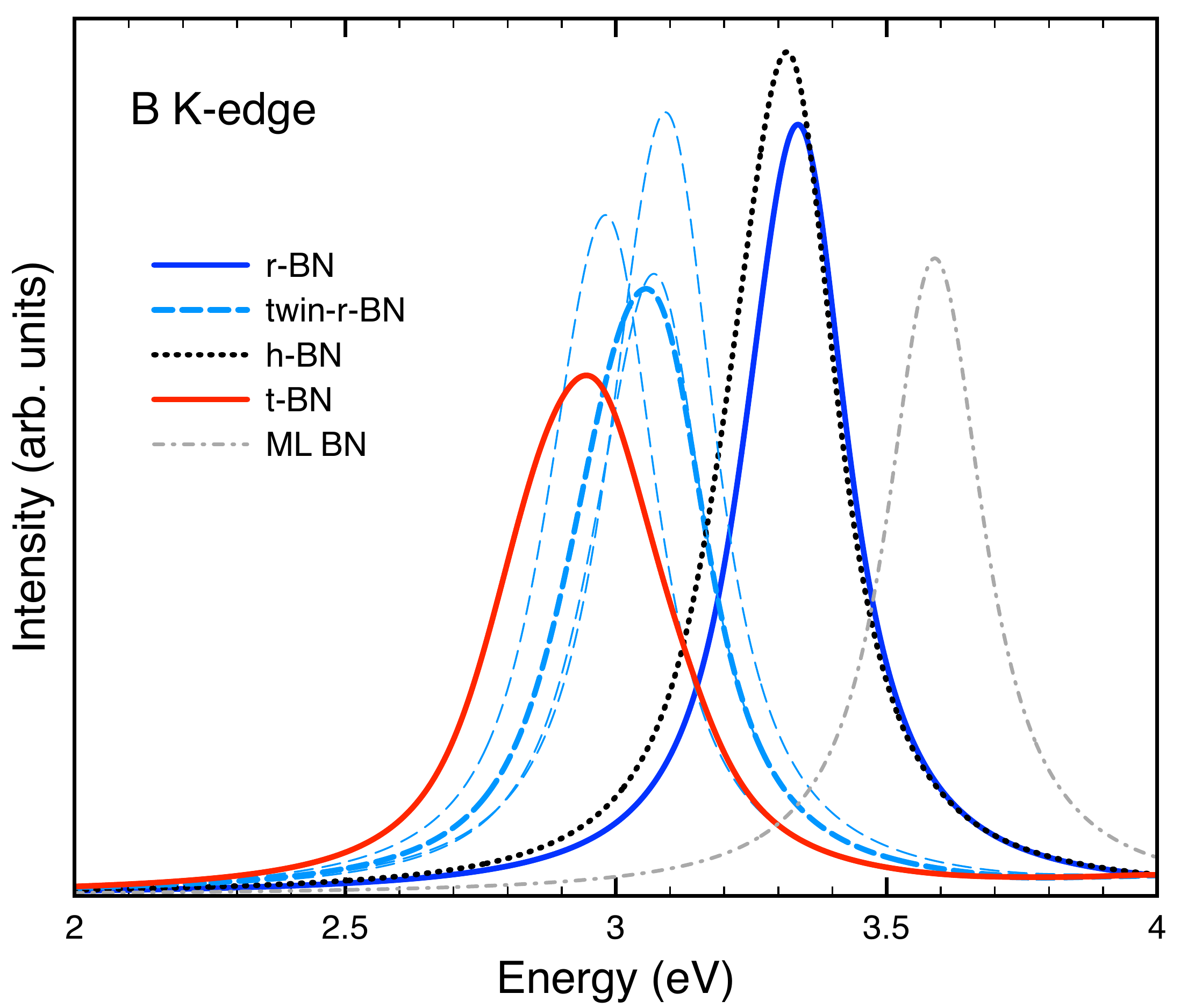} 
\vspace{0.2cm} 
\caption[] {Theoretical B K-edge XANES spectra of the rhombohedral (blue line), twinned rhombohedral (cyan dashed line), 
hexagonal (dotted line), turbostratic (red line), and ML (gray dot-dashed line) BN structures, focusing on the $\pi^*$ core-edge.}
\label{fig9}
\end{figure}

In Figure 9, a close-up of the calculated $\pi^*$ peaks at the B sites is shown, in order to focus on the chemical shifts originating from the different polytypes.
Here we will consider the chemical shift of $\pi^*$ with reference to the rhombohedral structure (blue line) at the higher energy. 
In all the studied sp$^2$ BN polytypes, the 2D hexagonal lattice planes are practically identical, with possibly a very small difference in the lattice parameter a.
The origin of the differences must then stem from the type of stacking of their respective planes and the interplanar distances.
The single monolayer BN system (gray dot-dashed line) marks the higher-energy limit for the shift of the $\pi^*$ peak, by increasing d in the other systems, at 0.25 eV above r-BN.
The turbostratic peak (red line) is represented by the average B1 spectra over the 10 different t-BN models, as seen in Fig.\ 8 a).
It is interesting to note that a smaller interlayer distance d as compared with the ordered phases, produces a more pronounced shift towards lower energies, while an increasingly larger distance induces a smaller shift.
We find that t-BN gives rise to the largest difference compared to r-BN, with a 0.39 eV shift towards lower energy.
The twinned r-BN model has three peaks (cyan thin long-dashed line) originating from inequivalent B atoms, resulting in an overall peak (cyan dashed line) shift of 0.28 eV.
h-BN (dotted line) in turn has the smallest shift, 0.02 eV.

The trend of the shifts roughly follows the calculated band gaps in Table 1, though with a larger difference between the t-BN and twin-r-BN models.
This difference indicates a sensitive dependence of the final state core-hole screening on the precise structure.
For the r-, h- and twin-r-BN systems, we checked the dependence of the $\pi^*$ peak position for larger supercells increasing to 4 x 4 planes (32 atoms), almost doubling the number of atoms. Only small differences were found compared with the smaller cells, with the $\pi^*$ energy shift between r-BN and h-BN and with twin-r-BN increasing by 0.01 eV, respectively.

A small splitting of the $\pi^*$ peak by 0.1 eV was found for h-BN in the experimental work by Li et al.~\cite{Li2012}. 
However, the large split of 0.7 eV as shown in the present work indicates a different origin. Previous calculations by McDougall et al.~\cite{McDougall2014} for h-BN, together with alternative variations in the stacking, only show small shifts of maximum 0.11 eV towards lower energy. Their largest shift was obtained in a model structure in which a layer slided along the basal plane such that the B atoms are on top, and the N atoms in-between, in comparison to the two hexagonal lattice planes in the unit cell.
In previous XANES measurements, a shift of $\sim$0.3 eV towards lower energy was found for the t-BN $\pi^*$ peak compared with h-BN~\cite{Zhou2006} and a shift of $\sim$0.1 eV between r- and h-BN was observed at the B K-edge~\cite{Terminello1994}.

By applying the BSE method within many-body perturbation theory, the strong $\pi^*$ resonance at the B K-edge of h-BN can be attributed to a strongly bound core-exciton state~\cite{Shirley2000, Olovsson2019}. It has also been shown that in order to reproduce the splitting into the $\sigma_1^*$ and $\sigma_2^*$ peaks of the "camel-back" feature, symmetry breaking due to vibrational effects needs to be included~\cite{Karsai2018,Olovsson2019}. 
While significantly affecting in-plane $\sigma$ bonds, a similar behaviour was not found for the $\pi^*$-edge. Thus, vibrational effects would lead to some broadening of the $\pi^*$-peak instead of a split at the B K-edge. For the corresponding N K-edge in h-BN, a small broadening was found~\cite{Vinson2017}.

The B K-edge is clearly sensitive to the precise stacking and orientation of the 2D hexagonal lattice sheets.
The larger variation in the stacking of planes, including translational randomness, gives a larger shift for the t-BN models as compared with r-BN.
Due to the sharp $\pi^*$ resonance at the B K-edge, it is feasible to distinguish between domains originating from different boron nitride structures in a sample.
In future studies it is of interest to consider how to improve the theoretical modelling to obtain more accurate chemical shifts. 
Here, one can for instance consider different theoretical methods, such as BSE, and more advanced modelling of t-BN structures.

\section{Conclusions}
The combination of high-resolution XANES measurements and {\it ab initio} calculations for the B and N K-edge are used to identify the constituent structures in sp$^2$ layered rhombohedral boron nitride. 
As shown in this work, the sharp $\pi^*$ XANES resonance at the B K-edge peak is highly sensitive to the precise stacking and orientation of the 2D hexagonal lattice sheets.
A splitting of the B 1s $\pi^*$ peak signifies defects or a multiphase structure of the material.
The large 0.7 eV $\pi^*$ split in the measurement is best matched in the calculations by a 0.39 eV shift obtained from turbostratic model structures, which for computational efficiency includes random translations but neglects random rotations.
This indicates that the composition of the investigated boron nitride sample is a superposition of two polytypes; an ordered rhombohedral crystal structure and a disordered turbostratic structure. 
In this way, it can be possible to determine the crystalline quality of rhombohedral and other boron nitride samples.
The shift of the $\pi^*$ peak is indicative of a sensitivity to the interplanar bond strength for different BN polytypes, also implying that the band gap can be tuned by the choice of polytypes.
We note that the combination of experiment and theory can be used as a fingerprint method for other 2D materials with sharp XANES features, such as a strongly bound core-exciton state.

\section{Acknowledgement}
We would like to thank Mikhail Chubarov and in particular the late Anne Henry for encouraging discussions and providing the r-BN sample. 
Research conducted at MAX IV, a Swedish national user facility, is supported by the Swedish Research council under contract 2018-07152, the Swedish Governmental Agency for Innovation Systems under contract 2018-04969, and Formas under contract 2019-02496. We thank the Swedish Government Strategic Research Area in Materials Science on Functional Materials at Link\"{o}ping University (Faculty Grant SFO-Mat-LiU No.\ 2009 00971). M.M. acknowledges financial support from the Swedish Energy Research (Grant No.\ 43606-1) and the Carl Tryggers Foundation (CTS20:272, CTS16:303, CTS14:310). The calculations were performed using supercomputer resources provided by the Swedish National Infrastructure for Computing (SNIC) at the National Supercomputer Centre (NSC), Link\"{o}ping University, partially funded by the Swedish Research Council through grant agreement no.\ 2018-05973.



\end{document}